\documentstyle[prl,aps,multicol]{revtex}
\input{epsf}

\newcommand{\kcl}{{\tilde k}}
\newcommand{\ti}{t'}

\newcommand{\pG}{{\bf p}}
\newcommand{\jG}{{\bf j}}
\newcommand{\Up}{{\hat{U}}}

\newcommand{\Hp}{{\hat{H}}}
\newcommand{\Np}{{\hat{N}}}

\begin{document}
\draft
\title{Stable Quantum Resonances in Atom Optics}

\author{Shmuel Fishman$^{1}$, Italo Guarneri$^{2,3,4}$, 
Laura Rebuzzini$^{2}$}
\address {$^1$ Physics Department, Technion, Haifa 32000, Israel}
\address{$^2$ Centro di Ricerca per i Sistemi Dinamici, 
Universit\`a dell'Insubria a Como, via Valleggio 11, 22100 Como, Italy}
\address{$^3$  Istituto Nazionale per la Fisica della Materia, 
via Celoria 16, 20133 Milano, Italy}
\address{$^4$ Istituto Nazionale di Fisica Nucleare, Sezione di Pavia, 
via Bassi 6, 27100 Pavia, Italy} 

\maketitle

\begin{abstract}
{A theory for stabilization of quantum resonances by a mechanism similar to one leading to classical
resonances in nonlinear systems is presented. It explains recent surprising experimental results,
obtained for cold Cesium atoms when driven in the presence of gravity, and leads to  
further predictions. The
theory makes use of invariance properties of the system, that are similar to those of solids, allowing 
for separation into independent kicked rotor problems.  The analysis relies on a fictitious 
classical limit where the small parameter is {\em not} Planck's constant, but rather the detuning from
the frequency that is resonant in absence of gravity.}
\end{abstract}                      
\pacs{PACS numbers: 05.45.Mt, 03.75.-b, 42.50.Vk} 

\begin{multicols}{2}
\narrowtext

Resonances are among the most typical manifestations of the quantum mechanical 
behavior. They are
characterized by a  precise relation between the parameters of the system 
and an external
driving; for instance, resonant excitation of an atom requires that the 
difference between two levels
be equal to the frequency of the driving field times Planck's constant. 
Classical resonances of
nonlinear systems are more robust. Regions of appreciable size in phase space may respond in a
resonant way, while other regions are chaotic. In recent experiments 
\cite{Ox1,Ox2,Ox3} it was found that a quantum system that is {\em far} from 
its semiclassical limit
exhibits resonant behavior very similar in nature to the one 
found for classical nonlinear
systems: it is robust, and the high precision that is characteristic of the usual quantum resonances
is not required. It is observed for various values of the parameters and of the initial conditions.
In this Letter the experimental results are explained and further predictions are made. 
Details are given in \cite{long}. 
The theory reveals how resonances, similar to those characteristic of classical systems, may  
arise in a quantum system, far from the semiclassical limit. The theoretical model of the
experiment is a variant of the kicked rotor, that is a standard system used in the investigation of
classical Hamiltonian chaos and its manifestations in quantum mechanical systems
\cite{KR,lichtenbergb}. In experimental realizations of the model, laser-cooled 
atoms are driven by application of a standing electromagnetic wave. 
The frequency of the wave is slightly detuned from resonance, so a
dipole moment is induced in the atom. This moment couples with the driving field, giving rise to a
net force on the center of mass of the atoms, proportional to the square of the electric field
\cite{GSZ,CT}. As the wave is periodic in space, the atom is thus subjected to a periodic potential.
The wave is turned on and off periodically in time, and the time it is on is much shorter then the
time it is off. A realization of a periodically kicked particle is then obtained.
For atoms driven in the horizontal direction, effects of gravity are negligible, and localization 
in momentum is observed\cite{raizen1}, as theoretically expected (see below). 

In the experiments that provide the subject of the
theoretical analysis of the present letter, laser cooled Cesium atoms were driven in the vertical
direction, and gravity was found to produce new  remarkable effects. Localization was  
destroyed; furthermore,  
in the vicinity of
the frequencies that are resonant in absence of gravity, a new type of ballistic spread in momentum
was experimentally observed \cite{Ox1,Ox2,Ox3}. A fraction of the atoms were steadily accelerated, at
a rate which was faster or slower than the gravitational acceleration depending on what side of the
resonance the driving frequency was. Such atoms were exempt from the diffusive spread that took place
for the other atoms, and their acceleration depended on the difference between the driving and the
resonant frequencies. 
The dynamics of the atoms that are falling as a result of gravity and are 
kicked by the
external field is modeled by the time-dependent Hamiltonian:
\begin{equation}
\label{ham3}
{\hat H}(\ti)\;=\;\frac{{\hat P}^2}{2}-\frac{\eta}{\tau}{\hat X}
+k\cos({\hat X})\sum\limits_{t=-\infty}^{+\infty}\delta(\ti-t\tau)\;,
\end{equation}
where $\ti$ is the continuous time variable, the integer 
variable $t$ counts the kicks, ${\hat P},{\hat X}$ are the momentum 
and the position operator respectively. Units are chosen so that 
the mass of the atoms is $1$, the Planck's constant is $1$, 
and the spatial period of the kicks is $2\pi$. The dimensionless 
parameters $k,\eta,\tau$ fully characterize the dynamics. 
They are expressed in terms of the physical parameters 
as follows: $k=\kappa/\hbar$, $\tau=\hbar TG^2/M$, $\eta=MgT/(\hbar
G)$, where $M,T,\kappa, g$ are the mass of the atoms, the kicking period, 
the kick strength, and the 
gravitational acceleration  respectively, and 
$2\pi/G$ is the spatial period of the kicks. The positive 
$x-$direction is that of the gravitational acceleration. 
{\it Throughout the following, time is a discrete 
variable, given by
the kick counter $t$.}

It is expedient  to measure the momentum in the 
free falling frame, notably to replace 
$\hat{P}-\frac{\eta}{\tau}\ti$ by $\hat{P}$, resulting in the 
time-dependent Hamiltonian :
\begin{equation}
\label{ham2}
{\hat H}(\ti)\;=\;
\frac{1}{2} (\hat{P}+\frac{\eta}{\tau} \ti)^2+k\cos (\hat{X})\sum\limits_
{t=-\infty}^{+\infty}
\delta(\ti-t\tau)\;.
\end{equation}
 
\begin{figure}
\centerline{\epsfxsize=7cm \epsfbox{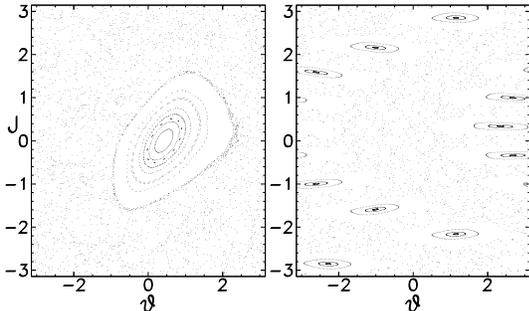}}
\caption{\small Phase portraits for the map (\ref{clmap1}) on the 2-torus,  
with $\eta/\tau=0.01579$. Left: stable (1,0) fixed point $J_0=0$, 
$\theta_0=0.42$, for $\tau=5.86, \kcl=1.329$. Right: stable 
(10,1) periodic orbit at $\tau=6.31, \kcl=0.067$.}
\label{fig1}
\end{figure}

This Hamiltonian 
is related to (\ref{ham3})
by the gauge transformation $e^{i\eta{\hat X}\ti/\tau}$.
The main advantage of (\ref{ham2}) compared to (\ref{ham3}) is that it depends on $\hat{X}$ only via 
$\cos(\hat{X})$. Consequently 
the evolution only mixes momenta which differ  
by integers: hence,  quasi-momentum is 
conserved. In absence of gravity ($\eta=0$), 
(\ref{ham2}) reduces to the Hamiltonian of the kicked rotor
(classically resulting in the Standard Map) with stochasticity parameter $K=k \tau$. For $K>K_c \approx 0.9716$
diffusion in momentum takes place. For values of $K$ near to integer multiples of $2\pi$
accelerator modes are found \cite{lichtenbergb}. These are stable phase space islands that
travel ballistically, resulting in linear, rather than diffusive growth of momentum with time. The 
quantum mechanical study of the systems described by (\ref{ham2}) starts with 
decomposing the momentum as $p=n+\beta$, where $n,\beta$ denote the integer and the fractional parts
of the momentum $p$ respectively. Since the quasimomentum $\beta$ is conserved only $n$ varies in
the course of the dynamics. The evolution is equivalent  to that of a superposition of
independent kicked rotors, each characterized by a different value of $\beta$.  Such a rotor will be
called $\beta$-rotor and the one step evolution operator is:
\begin{equation}
\label{Ufall}
\Up_{\beta}(t)\;=\;e^{-ik\cos({\hat\theta})}\;e^{-i\frac{\tau}{2} (\Np+
\beta+\eta
t+\eta/2)^2}.
\end{equation}
where $\theta=x$ mod$( 2 \pi)$ and the momentum operator is $\Np=-i\frac{d}{d\theta}$. For $\beta=0$
and $\eta=0$ this is the usual kicked rotor. The
classical diffusion is then suppressed by a mechanism that is similar to Anderson localization in
disordered solids \cite{KR,FGP}. For $\tau=2\pi l/m$, where $l$ and $m$
are integers, the eigenstates of (\ref{Ufall}) are extended in momentum
and  ballistic growth of momentum takes
place in most cases. This is the quantum resonance \cite{IzShep}. The Talbot length corresponds to $4 \pi$. 
For fixed $\beta \neq 0$ and  for typical irrational $\tau/2\pi$ the 
classical diffusion is 
again suppressed, as was observed in experiments on laser-cooled Sodium and Cesium atoms \cite{raizen1}. 
Quantum resonances only occur at special values of $\beta$, hence their experimental 
observation is difficult \cite{raizen2}, since it is impossible to 
prepare all
atoms with the same quasimomentum $\beta$. Averaging over $\beta$ results in linear (rather than  
quadratic) growth of the
squared momentum\cite{long}. This
demonstrates the high sensitivity of the quantum resonances to fine 
experimental details.

In the presence of gravity, $\eta \neq 0$, the evolution operator (\ref{Ufall}) is time
dependent. The localization in
momentum is destroyed, as it was experimentally observed \cite{Ox1,Ox2,Ox3}. 
While this is expected for time dependent random potentials, the present 
time dependence is just quasi-periodic, 
so a deeper analysis is required. However, 
the most surprising experimental result was 
that an appreciable fraction of atoms were found to accelerate (in the free falling
frame) for various values of the experimental parameters, for values of $\tau$ in intervals of
appreciable size around integer multiples of $2\pi$.  Here is  a quantum resonance that is
robust, in contrast to the usual quantum resonances. It is reminiscent of classical nonlinear
resonances, namely the accelerator modes of the Standard Map; however, 
{\it it has no counterpart 
in the classical limit of (\ref{ham2})}. In what follows a theoretical
explanation of this effect is presented \cite{long}.

We consider the case when  $\tau$ is close to a resonant 
value $2\pi l$ ($l>0$ integer), and the kicking strength $k$ 
is large. We hence write $\tau=2\pi l+\epsilon$, $k={\tilde k}/|\epsilon|$ 
with $\epsilon$ small.  
Noting $e^{-i\pi ln^2}=e^{-i\pi ln}$, (\ref{Ufall}) takes the form 
(apart from an irrelevant  phase factor):
\begin{equation}
\label{epsdin1}
\Up_{\beta}(t)\;=\;e^{-\frac{i}{|\epsilon|}{\tilde k}\cos{\hat \theta}}\;
e^{-\frac{i}{|\epsilon|}\Hp_{\beta}({\hat I},t)}
\end{equation}
where
\begin{equation}
\label{epsdin2}
\begin{array}{l}
{\hat I}=|\epsilon|\Np=-i|\epsilon|\frac{d}{d\theta}\\
\Hp_{\beta}({\hat I},t)=\frac12\; \mbox{\rm sign}(\epsilon) {\hat I}^2+
{\hat I}(\pi l+\tau (\beta+t\eta+\frac{\eta}{2}))\\
\end{array}
\end{equation}
If $|\epsilon|$ is assigned the role of Planck's constant, then 
 (\ref{epsdin1},\ref{epsdin2}) is the formal quantization 
of either of the  following classical (time-dependent) maps :
\begin{equation}
\label{clmap}
\begin{array}{l}
I_{t+1}=I_{t}+{\tilde k}\sin (\theta_{t+1})\\
\theta_{t+1}=\theta_t\pm I_t+\pi l+\tau(\beta+t\eta+\eta/2)\;\mbox{\rm mod}
(2\pi), \\
\end{array}
\end{equation}
where $\pm$ has to be chosen according to the sign of $\epsilon$.
The small $|\epsilon|$ asymptotics of the quantum 
$\beta-$ rotor is thus equivalent to a  
quasi-classical approximation based on the ``classical'' 
dynamics  (\ref{clmap}). 
We emphasize that ``classical'' here is not 
related to the $\hbar\to 0$ limit but to the limit $\epsilon\to 0$ instead.
The two limits are actually incompatible with each other except possibly 
when $l=0$. For the sake of clarity the term ``$\epsilon-$classical'' will
be used in the following.

Changing variable 
to $J_t=I_t\pm \pi l\pm\tau (\beta+\eta t +\eta/2)$ removes the explicit 
time dependence of the maps (\ref{clmap}), yielding:
\begin{equation}
\label{clmap1}
J_{t+1}=J_t+{\tilde k}\sin (\theta_{t+1})\pm\tau\eta\;,\;\;
\;\theta_{t+1}=\theta_t\pm J_t\;.
\end{equation}
If $J,\theta$ are taken mod$(2\pi)$, then (\ref{clmap1}) define 
maps of the $2-$torus in itself.  
Let $J_0,\theta_0$ be a
period $\pG$-fixed point  of either of the toral maps thus defined. 
Then  
iteration of (\ref{clmap1}) yields, at $t=\pG$ :  
\begin{equation}
\label{fixp} 
J_{\pG}=J_0+2\pi \jG\;\;,\;\;\theta_{\pG}=\theta_0+2\pi m
\end{equation} 
for some integers $\jG\;,\;m$. In terms of the original 
dynamics (\ref{clmap}) this yields a family of orbits 
such that, for all integer $t$, 
\begin{equation}
\label{growth}
\theta_{\pG t}=\theta_0=\vartheta_0\;\mbox{\rm mod}(2\pi)\;\;,\;\;
I_{t\pG}=I_0+a\pG t\;, 
\end{equation}
where $a=\mp \tau\eta+2\pi\jG/\pG$ and 
\begin{equation}
\label{accmod1}
I_0=J_0\mp\pi l\mp\tau(\beta+\eta/2)+2\pi m'\;\;,
\end{equation}
with $m'$ any integer. Thus primitive periodic orbits of 
the toral maps (\ref{clmap1}) correspond to families of 
{\it accelerator orbits} of the dynamics (\ref{clmap}), marked 
by linear average growth of momentum with time. 
If the  periodic orbits are stable, 
then the accelerator orbits are surrounded by 
islands  of positive measure in phase space, also leading to ballistic 
(linear) average growth  of momentum in time. These are named 
{\it accelerator modes}. They are characterized by the 
integer couple $(\pG,\jG)$ formed by the {\it order} $\pG$ 
and by the {\it jumping index} $\jG$.

\begin{figure}
\centerline{\epsfxsize=7cm\epsfbox{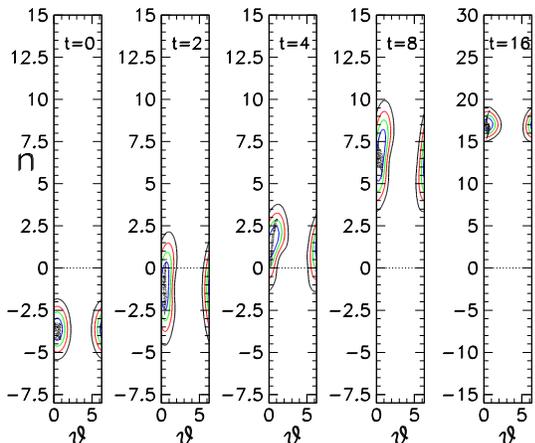}}
\caption{\small Contour plots of Husimi functions at times  
$t=$$0$,$2$,$4$,$8$,$16$ 
for a $\beta-$rotor  
initially prepared in the  coherent state centered 
at the $\epsilon-$classical $(1,0)$ accelerator mode 
of Fig.1 (left).
The black spots in the centers of the contours are an ensemble of 
classical points of size  $\hbar=1$, evolving according to 
the $\epsilon-$classical dynamics (\ref{clmap}).}
\label{fig2}
\end{figure}

Modes of order $1$  correspond to  
fixed points of (\ref{clmap1}). They are given on the 2-torus 
by $J_0=0$, $\theta_0=\theta_{\jG}$, where
\begin{equation}
\label{fix}
\sin(\theta_{\jG})\;=\;(2\pi\jG\mp\tau\eta)/\kcl
\end{equation}
and $\jG$ is any integer such that the rhs is not larger than 
$1$ in absolute value.
A linear stability analysis  shows that 
 for any 
integer $\jG$ each map (\ref{clmap1}) has exactly one 
stable period-$1$ fixed point on the 2-torus,  
given by (\ref{fix}) if, and only if,
\begin{equation}
\label{stabcond}
|2\pi\jG\mp\tau\eta|<\kcl<\sqrt{16+(2\pi\jG\mp\tau\eta)^2}
\end{equation}
When  $\kcl$ increases beyond the right-hand limit, 
such  fixed points turn unstable and   
bifurcations occur. At small values of $\kcl$ higher-order accelerator modes
appear, associated with higher-period primary orbits. 
Examples of stable periodic orbits  near $\tau=2 \pi$ are presented in Fig.1. 

\noindent Initial physical momenta $n_0=|\epsilon|^{-1}I_0$ for
$\epsilon-$classical accelerator modes are obtained from (\ref{accmod1}) 
for any $0\leq \beta<1$.
If the stable 
islands associated with $\epsilon-$classical accelerator modes have a large 
area compared to $|\epsilon|$, then they 
 may trap some of the rotor's wave packet and give rise to 
{\it quantum accelerator modes}  traveling  in physical momentum 
space with speed $\sim a/|\epsilon|=-\tau\eta/\epsilon+2\pi\jG/(\pG |\epsilon|)$.
In order that such modes
may be observed, the phase space distribution associated with the initial 
rotor state must significantly overlap the islands. 
This picture is confirmed by numerical simulations. Fig.2 
 shows the quantum 
phase-space evolution  of a $\beta-$rotor with $\beta=0.2188$  
started in the  coherent state centered at the position of a  
$(1,0)$- accelerator mode. The Husimi functions  
computed at subsequent times  closely  follow the motion of the 
$\epsilon-$classical mode. 
In the $\epsilon-$ semiclassical regime, accelerator modes are expected to decay
exponentially in time 
due to quantum tunneling out 
of the classical islands, with  decay rate   
$\gamma_{\epsilon}\propto \exp{-(\mbox{\rm const.}/|\epsilon|)}$. 
This decay was 
also numerically confirmed \cite{long}. 

In the experiments described in \cite{Ox1,Ox2,Ox3}, the initial state of the falling atoms 
is satisfactorily described by 
an incoherent Gaussian mixture of momentum eigenstates  centered at $p=0$, with rms deviation $\simeq
2.55$ in our units. This is equivalent to a statistical ensemble 
of $\beta$-rotors.  
The above theory describes this situation for small $\epsilon$, where the $\epsilon$-
classical approximation holds. From (\ref{growth}) it is found that the $\epsilon$-classical
$(\pG,\jG)$-mode started at $t=0$ with $I_0=n_0|\epsilon|$ is located at time $t$ at the momentum:
\begin{equation}
\label{curv}
n\simeq n_0- t\tau\eta/\epsilon+2\pi t \jG/(\pG |\epsilon|)\;.
\end{equation}
In Fig.3 this prediction is compared to the results of numerical
simulations in the vicinity of $\tau=2 \pi$ for $k=0.8\pi$ and $\eta/\tau=0.01579$ (that is the
value of the gravitational acceleration in our units). The time is $t=60$ kicks. The initial
state is an incoherent mixture of $50$ momentum states, distributed as reported in experimental
papers. Lines are as  predicted by (\ref{curv}). Good agreement is found. The mode $(1,0)$ was
identified experimentally (as can be seen on comparing  Fig.3 to Fig.2 
of \cite{Ox1}; also 
recall that our positive direction is that of the gravitational acceleration). 
Further higher order modes are predicted in the present work. Longer time is required for their resolution, 
because they move slower, yet they may be eventually resolved, see  
the inset of Fig.3.  
Similar behavior is predicted for $\tau \approx 4 \pi$ (as described in \cite{long}),  and for
higher multiples of $2\pi$ as well. 
We finally remark that (i) quantization of the 
$\beta-$ rotors momentum enhances quantum modes at discrete values of 
$\beta$, given by (\ref{accmod1}) with $I_0$ an integer 
multiple of $\epsilon$. This results in preferred values for 
physical momentum $p=n+\beta$, spaced by $\approx 2\pi l/\tau$. 
This was experimentally detected, and it is seen in the fine 
structure of the modes in Fig.3; (ii)  
the accelerator modes of the $\beta$-rotors result in acceleration 
of the particle also in the case when the initial state is a coherent superposition 
of momentum eigenstates\cite{long}.

\begin{figure}
\centerline{\epsfxsize=6.5cm\epsfbox{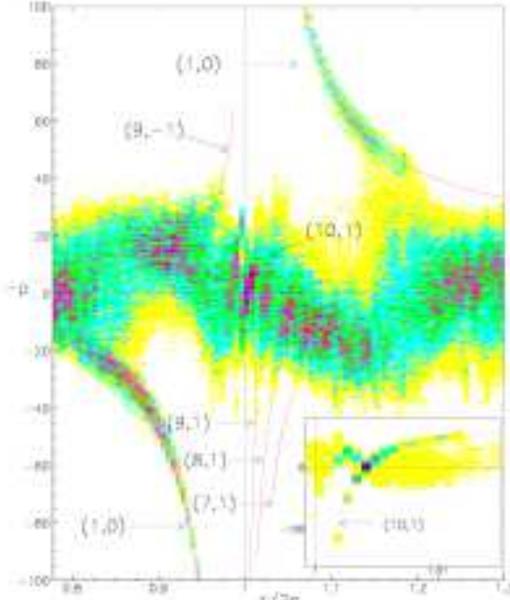}}
\caption{\small Momentum distribution in the falling frame 
at $t=60$ kicks for different values of the kicking period $\tau$ 
near $\tau=2\pi$. Note the negative sign of $p$. 
Darker regions correspond to higher 
probability. The initial state is  a mixture of plane waves sampled 
from a gaussian distribution of momenta. Full lines are the 
theoretical curves (\ref{curv}), with 
orders and  jumping indexes as  
indicated  by the arrows. The inset shows data at $t=400$.}
\label{fig3}
\end{figure}

The key steps in the theory presented in this work are:
({\it i}) Transformation to the free falling frame leading 
to the Hamiltonian (\ref{ham2});
({\it ii}) Conservation of the quasimomentum $\beta$, allowing for 
introduction of the $\beta$-rotors;
({\it iii}) Approximation of the motion by the one generated by the
$\epsilon$-classical Hamiltonian 
(\ref{epsdin2}) near $\tau=2\pi l$;
({\it iv}) $\epsilon$-classical accelerator modes (\ref{accmod1}) 
dominate the dynamics near $\tau=2\pi l$. 

\noindent
What stabilizes these accelerator modes? 
On top of the long wavelength behavior related to the resonance
$\tau=2\pi l$, there is a short wavelength behavior. Its 
dynamics is dominated by the fictitious classical
mechanics termed $\epsilon$-classical. This is reminiscent 
of  the ``caustics without rays''
\cite{berry}.  The generality of this mechanism for nonlinear
stabilization of quantum resonances should be explored. The fact that 
the intermediate-time dynamics is
dominated by a discrete set of modes, which exponentially decay in time, 
bears a distinct resemblance to
the Wannier-Stark problem of a Bloch particle in a constant field \cite{WS}. 
How far this analogy
carries is, in our opinion, an interesting theoretical question.
Other theoretical questions well worth further analysis are 
about the nature of the 
asymptotic (in time) behavior (in particular, the mechanism for 
destruction of localization), 
and  the nature of the $\epsilon$-semiclassical systematic expansion.
On the experimental side, it would be interesting to classify  the various 
accelerator modes, namely to perform
``accelerator mode spectroscopy''; and also  to experimentally explore 
differences in the quantum dynamics between coherent and incoherent 
superpositions in the initial state
of the atoms. This may be of importance for atomic interferometry and for quantum computation.

\noindent{\bf Acknowledgments.}
This research was supported in part by
{\it PRIN-2000: Chaos and localisation in classical and quantum mechanics},
by the US-Israel Binational Science
Foundation
(BSF), by the US National Science Foundation under Grant No. PHY99-07949,
by the
Minerva Center of Nonlinear Physics of Complex Systems,
by the Max Planck Institute for
the Physics of Complex Systems in Dresden,
and by the fund for Promotion of Research
at the Technion. Useful discussions with
M.Raizen, M.V. Berry, K.Burnett, M. d'Arcy, M. Oberthaler, R. Godun, 
Y.Gefen, and S.Wimberger  are acknowledged.

\end{multicols}

\begin{thebibliography}{99}


\bibitem{Ox1} M.K. Oberthaler, R.M.Godun, M.B. d'Arcy, G.S. Summy,
and K. Burnett, Phys. Rev. Lett. {\bf 83}, 4447 (1999).

\bibitem{Ox2} R.M.Godun, M.B. d'Arcy, M.K. Oberthaler, G.S. Summy,
and K. Burnett, Phys. Rev. A{\bf 62}, 013411 (2000).


\bibitem{Ox3}  M.B. d'Arcy, R.M.Godun, M.K. Oberthaler, G.S. Summy,
and K. Burnett, S.A. Gardiner, Phys. Rev. {\bf E 64}, 056233 (2001);
M.B. d'Arcy, R.M.Godun, M.K. Oberthaler, D. Cassettari and G.S. Summy,
Phys. Rev. Lett.  {\bf 87}, 074102 (2001).

\bibitem{long} S. Fishman, I. Guarneri and L. Rebuzzini, {\it 
A Theory for Quantum Accelerator Modes in Atom Optics}, To be 
published in J. Stat. Phys. (nlin.CD/0202047)


\bibitem{KR} for reviews see, e.g.: F.M. Izrailev, Phys. Rep. {\bf 196}, 299
(1991); S. Fishman, in {\it Proceedings of the International School of Physics
Enrico Fermi: Varenna Course CXIX}, G.Casati, I. Guarneri and U.Smilansky
eds., North Holland 1993, p.187.


\bibitem{lichtenbergb}
A.J. Lichtenberg and M.A. Lieberman, {\em Regular and Chaotic Dynamics}, (Springer-Verlag, NY, 1992)

\bibitem{GSZ}
R. Graham, M. Schlautmann and P. Zoller, Phys. Rev. {\bf A 45}, R19 (1992).

\bibitem{CT}
C. Cohen-Tannoudji, J. Dupont-Roc, G. Grynberg,
{\it Atom-Photon Interactions, Basic Processes and Application}, (Wiley, NY
1992), P. 456 

\bibitem{raizen1}
D.A. Steck, V. Milner, W.H. Oskay, and M.G. Raizen,
Phys. Rev. {\bf E 62}, 3461 (2000);
F.~L. Moore, J.~C. Robinson, C.~F. Bharucha, Bala Sundaram, and
M.~G. Raizen, Phys. Rev. Lett. {\bf 75}, 4598 (1995);
C.~F. Bharucha, J.~C. Robinson, F.~L. Moore, Qian Niu, Bala Sundaram, and
M.~G. Raizen, Phys. Rev. {\bf E 60}, 3881 (1999);
B.G. Klappauf, W.H. Oskay, D.A. Steck amd M.G. Raizen, Physica
(Amsterdam) {\bf 131 D}, 78 (1999).

\bibitem{FGP} S. Fishman, D.R. Grempel, and R.E. Prange, Phys. Rev. Lett.
{\bf 49}, 509 (1982); D.R. Grempel, R.E. Prange, and S. Fishman, Phys.
Rev. A {\bf 29}, 1639 (1984).

\bibitem{IzShep} F.M.Izrailev and D.L.Shepelyansky, Sov. Phys. Dokl.
{\bf 24}, 996 (1979); G.Casati and I.Guarneri, Comm. Math. Phys.
 {\bf 95}, 121 (1984).


\bibitem{raizen2}
W.~H. Oskay, D.~A. Steck, V. Milner, B.~G. Klappauf, and M.~G. Raizen,
Opt. Comm. {\bf 179}, 137 (2000).

\bibitem{berry} M.V. Berry and E. Bodenschatz, Jour. of Modern Optics, {\bf 46}, 349 (1999).

\bibitem{WS} for a review see, e.g., G. Nenciu,
Rev. Mod. Phys. {\bf 63}, 91 (1991) and references therein.







\end{thebibliography}
\end{document}